\documentclass[12pt]{article}
\usepackage{graphicx}
\usepackage{color}
\usepackage{amsmath,slashed, amssymb}
\usepackage{fancyhdr}

\def\beq{\begin{eqnarray}}
\def\eeq{\end{eqnarray}}
\def\bea{\begin{eqnarray*}}
\def\eea{\end{eqnarray*}}
\def\bal{\begin{align}}
\def\eal{\end{align}}

\def\centeron#1#2{{\setbox0=\hbox{#1}\setbox1=\hbox{#2}\ifdim
\wd1>\wd0\kern.5\wd1\kern-.5\wd0\fi
\copy0\kern-.5\wd0\kern-.5\wd1\copy1\ifdim\wd0>\wd1
\kern.5\wd0\kern-.5\wd1\fi}}
\def\ltap{\;\centeron{\raise.35ex\hbox{$<$}}{\lower.65ex\hbox{$\sim$}}\;}
\def\gtap{\;\centeron{\raise.35ex\hbox{$>$}}{\lower.65ex\hbox{$\sim$}}\;}

\def\singleandthirdspaced{\baselineskip=\normalbaselineskip\multiply
    \baselineskip by 130\divide\baselineskip by 100}

\newcommand{\newc}{\newcommand}
\newc{\qbar}{{\overline q}}
\newc{\Kahler}{K\"ahler }
\newc{\deltaGS}{\delta_{\rm GS}}

\begin{document}
\begin{titlepage}
\begin{flushright}
{\large hep-th/yymmnnn \\
SCIPP 14/11\\
}
\end{flushright}

\vskip 1.2cm

\begin{center}

{\LARGE\bf Hybrid Inflation with Planck Scale Fields}

\vskip 1.4cm

{\large Michael Dine and Laurel Stephenson-Haskins}
\\
\vskip 0.4cm
{\it Santa Cruz Institute for Particle Physics and
\\ Department of Physics,
     Santa Cruz CA 95064  } \\

\vskip 4pt

\vskip 1.5cm

\begin{abstract}
Observable B-mode polarization in the CMBR would point to a high scale of inflation and large field excursions during the inflationary era.  Non-compact string moduli spaces
are a suggestive setting for these phenomena.   
Although they are unlikely to be described by weak coupling models, effective field theories compatible with known features of cosmology do exist.
These models can be viewed as generalizations to a large field regime
of hybrid inflation.  We note close parallels to small and large field axion models.  This paper outlines the requirements for successful modular inflation, and gives examples of effective field theories
which satisfy them.  The required tunings are readily characterized.  These models can also be thought of as models of chaotic inflation, in a way we describe.
In the modular framework, one would expect that any would-be Peccei-Quinn symmetry would likely be badly broken
during inflation, and the axion would have Hubble scale mass; in this situation, isocurvature fluctuations would be suppressed and the initial misalignment angle would be fixed, rather than
being a random variable.  \end{abstract}

\end{center}

\vskip 1.0 cm

\end{titlepage}
\setcounter{footnote}{0} \setcounter{page}{2}
\setcounter{section}{0} \setcounter{subsection}{0}
\setcounter{subsubsection}{0}

\singleandthirdspaced

\section{Small vs. Large Field Inflation}

Models of slow roll inflation can be divided into two broad categories:  small field and large field, where the small or large is relative to the Planck scale, $M_p$ (there are many good reviews; on this point, see, for example, \cite{lyth}).  These two classes of theories differ dramatically in whether or not they predict observable gravity waves.  Each class of models poses theoretical challenges as well.  In large field models, if $\phi$ is the inflaton, the field responsible for inflation, then one can't analyze an effective action for $\phi$ in powers of $\phi/M_p$.  It is not, in fact, quite clear in what framework (outside of some larger theory of quantum gravity) one might understand such theories.  As we will review, small field models also cannot be completely understood without a complete underlying theory of gravity.  That said, the problem of inflation in these theories can be described by a small number of parameters.

The BICEP2 announcement of the possible observation of gravity waves in the CMB\cite{bicep2} brought the question of large vs small field inflation to the forefront.  While there is no longer any claim to an observation\cite{bicepplanck}, there are intense efforts to further constrain (or observe) B mode
polarization in the CMBR.
The BICEP2 result was suggestive of an energy sale of inflation
would be about $2 \times 10^{16}$ GeV; Planck set limits of order 1/2 of this\cite{planckinflation}.

A great deal has been written on the subject of large field inflation, trying to accommodate the original BICEP2 claim,
and suggesting, in any case, that such radiation should be observable.   This work can again be divided into two broad categories (with some overlap):  natural inflation\cite{natural} and chaotic inflation\cite{chaotic}.  Natural inflation involves axion-like fields, with decay constants larger than $M_p$.  Because such decay constants seem hard to realize in string theory\cite{banksdinefoxgorbatov}, much work has focussed on {\it monodromy inflation} and its variants (though see  \cite{choinatural}), in which axions transit many times their nominal periods\cite{monodromy}, or theories with multiple axions (or fields which can wander circuitously through field space)\cite{multipleaxions}.  Chaotic inflation involves fields with monomial potentials with very small coefficients.  As implemented in \cite{monodromy}, monodromy inflation is actually a realization of chaotic inflation, with a monomial potential for the inflaton.  It is argued that the features of the inflaton potential, in this case, can be understood within an ultraviolet complete theory, string theory.  Related ideas for achieving inflation have been considered in \cite{new1,new2,new3}.

In this note, we examine a different arena for inflation:  non-compact string moduli spaces.  Classically, string compactifications with zero cosmological constant (c.c.) typically exhibit moduli of various sorts.  Such light fields might exist quantum mechanically.  One possible explanation for this is low energy supersymmetry, where the light non-compact moduli would be superpartners of axions.   We will take this as our working model throughout this paper\footnote{An example of a landscape of non-supersymmetric vacua, in many cases with light moduli, appears in \cite{silversteinhandles}; there are likely other non-supersymmetric possibilities, including simply anthropic selection.}.    By low we mean that during inflation, the soft breaking terms, while possibly quite large compared to the scale at which supersymmetry is ultimately broken, are well below the energy scale of inflation.  Supersymmetry breaking during inflation has been discussed in\cite{dinerandallthomas}, where it is stressed that, for slowly varying fields,  it is possible to use a supersymmetric effective action to describe the dynamics of the inflaton.  This action can be organized in terms of a superpotential, Kahler potential, and higher derivative operators. The natural mass (curvature) scale for the moduli potential is of order $H_I$, the Hubble constant during inflation.  Given that slow-roll inflation requires an inflaton mass significantly less than $H_I$, the models we consider, like essentially all models of inflation, will require some level of tuning of parameters.  The inflaton may be part of a chiral multiplet responsible for supersymmetry breaking during inflation, i.e. its fermionic partner may be an approximate goldstino.  This multiplet need not not necessarily contain the Goldstino (longitudinal component of the gravitino) responsible for supersymmetry breaking at lower scales (we will not make any particular assumption about the final scale of supersymmetry breaking, other than that it is low compared to the scale of inflation).  

Ours is certainly not the first work to consider moduli as candidate inflatons.  Some discussion occurs already in \cite{banksmodulicosmology}.  More recently, models relying on quite specific features of particular string models have been considered in some detail\cite{quevedo1,burgesslargefield}.  Our goal here is to delineate some general issues.  Some of the models have features which correspond to some of the general features discussed here, but we do not believe that all have been treated before.

 We begin, in section \ref{smallfieldhybrid} with a brief review of conventional small field inflation models, and in particular of hybrid models.  The essence of such models is that inflation occurs on a pseudomoduli space\cite{dinepack}.  Standard analyses are restricted to small field inflation, and we ask what are the features intrinsic to such a regime.  We also review the challenges of understanding the spectral index in these theories.

We then make a short excursion into axion physics in section \ref{smallandlargeaxions}.  We note that there is a remarkably close parallel between large and small field inflation and large and small field solutions to the strong CP problem.  The large and small fields we refer to here are precisely the non-compact (pseudo) moduli accompanying the axions (saxions), and large and small, again, means relative to the Planck scale.  Indeed, existing small field solutions to the strong CP problem are quite complex and suffer from a lack of plausibility.  Large field solutions (implemented in string theory) require far less theoretical gymnastics.  They involve, however, regimes which are inherently inaccessible to weak coupling methods.  Indeed, there is tension between the requirement of small exponentials (large values of non-compact moduli) and the absence of small parameters in the theory\cite{dineseibergstrongcoupling}.  Within our present understanding, the large-field axion solution is at best a plausible hypothesis.

We argue that there are clear lessons from axion physics for inflation, and turn to modular inflation in section \ref{moduli}.  We will assume some 
We begin by noting that the non-compact moduli of string theory models have properties that would seem well suited to inflation.  We have in mind situations with at least some approximate supersymmetry, so that these moduli are accompanied by compact moduli (axions).  Such moduli naturally have Planck scale variations. At the same time, for large values of the moduli, the energy scale
is low compared to the Planck scale; in other words, the requirement of very small couplings is replaced by
a condition on the moduli fields.
We will argue that if inflation occurs in such a region of the (pseudo) moduli space, it can be described in the language of effective field theory, and one can formulate the required properties of that theory.  We will describe simple model field theories which are consistent with the data on inflation.  We stress, however, that these are only models.  Extracting such a structure from an underlying theory is beyond present theoretical technology.  The degree of (apparent) tuning in these models is readily characterized.  We will see that, in such a framework, the lower the scale of inflation, the greater the degree of tuning.  

Inflation on string moduli spaces provides a setting to address a much discussed issue:  the seeming incompatability of large field inflation and the axion solution of the strong CP problem\cite{thomasaxion}.   We recall that in such a framework, the Peccei-Quinn symmetry is necessarily an accident of the features of moduli fixing (and in particular the fixing of the saxion).  These features need not hold during inflation, so the Peccei-Quinn symmetry may be badly broken during this period, and axion fluctuations suppressed\cite{dineanisimov}.  We discuss some of the features required of the effective field theory
to achieve this.

We do not, of course, have a detailed string construction which we can understand at the level required to test this picture.  Instead, in the rest of this note, we describe the conditions on the field content and effective action of such theories required to obtain suitable inflation, consistent with current observations, including recent Planck and a possible observation of gravitational radiation.  We will then include a requirement that the universe, in its present stage, include an axion, with suitable properties to solve the strong CP problem, and behavior in the early universe which suppresses would-be isocurvature fluctuations.  
Some aspects of inflation on non-compact moduli spaces have been considered in \cite{burgesslargefield}, though with a somewhat different focus.

\section{Hybrid Inflation:  Small Field and Large Field}
\label{smallfieldhybrid}

Most models of large scale inflation are based either on polynomial potentials (``chaotic inflation") or axionic shift symmetries (``natural" inflation, and its variants including multi-axions and monodromy).  (For a recent attempts to reconcile small field models with a large scale of inflation, see \cite{rabycarpenter,hertzbergwilczek}).  But there is another possibility, raised by experience in supersymmetric field theories and string theory, involving non-compact moduli.

Indeed, one interesting class of inflationary models are so-called hybrid models\cite{lindehybrid,hybrid1,hybrid2,hybrid3,hybrid4}.   These are often described in terms of fields and potentials with rather detailed, special features, but in \cite{dinepack,dinehybrid2}, hybrid inflation was characterized in a more conceptual way.   Inflation is active
 all such models on a pseudomoduli space, in a region where superymmetry is badly broken (i.e. broken by a larger amount than in the present universe) and the potential is slowly varying; in fact, this is the defining feature of these theories.  For example, much has been made in these models of the role of a ``waterfall field", but in practice, inflation often ends well before reaching the regime where this field is active.   All that is really required is that the fields settle into a region with
 much smaller cosmological constant after inflation ends.

Essentially all hybrid models in the literature are small field models; this allows quite explicit constructions using rules of conventional effective field theory, but it is not clear that small field inflation is selected by any deeper principle.  The goal of this paper is to consider large field models.  This is made all the more important given
that the small field models would be ruled out by
observation of B-mode polarization (many were already ruled out by the Planck measurement of the spectral index, $n_s$\cite{planckinflation}, but for surviving alternatives, see, for example, \cite{pallisshafi,hilltop,dinehybrid2}).

It is worth reviewing the simplest small-field hybrid model.  Such a model is supersymmetric (this allows the natural appearance of a classical pseudomoduli space), with two fields, $I$ (which will play the role of the inflaton) and $\phi$ (usually referred to as the ``waterfall field."  The superpotential is taken to be:
\beq
W = I (\kappa \phi^2 - \mu^2)
\eeq
Classically, for large $I$, the potential is independent of $I$; the quantum mechanical corrections control the potential.  $\kappa$ is constrained to be extremely small in order that the fluctuation spectrum be of the correct size; $\kappa$ is proportional, in fact, to $V_I$, the energy during inflation.   The quantum corrections determine the slow roll parameters.

One expects corrections at least in powers of $M_p$.  If $I \ll M_p$, one can organize the effective field theory in powers of $I$.  Particularly critical are higher powers of $I$ in the \Kahler potential.  The quartic term in $K$,
\beq  
K = {\alpha \over M_p^2 }I^\dagger I I^\dagger I
\eeq
gives too large an $\eta$, for example, unless $\alpha$ is suitably small\cite{linderiotto,dinepack}.

It is also necessary to suppress high powers of $I$ in the superpotential.  In general, terms of the form
\beq
\delta W = {I^n \over M_p^{n-3}}
\eeq
will be allowed, and at least the low $n$ terms must be suppressed.
This might occur as a result of discrete symmetries.  The leading power of $I$ in the superpotential controls, for example, the scale of inflation. Higher powers allow larger scales; a scale of $10^{15}$ requires $n \ge 11$\cite{dinepack}.  In \cite{dinehybrid2}, it was argued, based on a systematic study of the effective action, that obtaining $n_s < 1$, consistent with Planck, required a balancing of \Kahler and superpotential corrections.

In small field inflation models, it is still necessary to have control over Planck scale corrections, and tuning of parameters (at least at the part in $10^{-2}$ level) seems required.  One needs a large discrete symmetry to account for a low scale of inflation,
and a very small dimensionless parameter, progressively smaller as the scale of inflation beomes smaller.   So the theoretical arguments for small field over large field inflation are hardly so persuasive.  
Given both the theoretical situation and the ongoing searches for tensor modes, it is clearly interesting to explore the possibility of inflation on moduli spaces with Planck scale fields undergoing variations of order Planck scale or larger.  Such moduli spaces are quite familiar from string theory.  Though a reliable construction may be difficult, we can look to such theories for insights into how such a theory of inflation might look.

\section{Small Field and Large Field Solutions to the Strong CP Problem}
\label{smallandlargeaxions}

 The situation in small vs large field inflation is reminiscent -- and as we will see closely related to -- that of the QCD axion.  To {\it solve} the strong CP problem, it is not enough to postulate the existence of a light pseudoscalar; one must account for an accidental global symmetry which is of extremely high {\it quality}\cite{dineetalquality}.  Here, too, there are small field and large field solutions.  Most models designed to obtain a Peccei-Quinn symmetry are constructed with small axion decay constant, $f_a \ll M_p$ (analogous to the small value of $I$).  In such constructions, $f_a$ is related to the expectation value of some field, $\phi$ (possibly a fundamental scalar or a composite operator). In this case, near the scale $f_a$ one can write an effective field theory, organized in powers of $\phi/M_p$, where we have taken the relevant scale to be $M_p$.  If the scale is lower, the problem of obtaining a suitable Peccei-Quinn symmetry, i.e.\! a solution of sufficient {\it quality}\cite{dineetalquality}, is more daunting.  It has long been appreciated that it is necessary to suppress many operators in order to have a sufficiently light axion to solve the strong CP problem.  This can be achieved through discrete symmetries, but the symmetries must be quite intricate.  For example, with a single $Z_N$, with supersymmetry one needs minimally $N=11$ or $N=12$; without supersymmetry $N$ must be somewhat larger\cite{lazaridesaxions,dinez12}.

String theory has long suggested a different, large field, perspective on the axion problem\cite{wittenaxion}.   This is particularly easy to describe
in cases where there is some approximate supersymmetry (unbroken supersymmetry at scales at least somewhat below the
string scale). In any theory with approximate supersymmetry, axions are accompanied by (non-compact) moduli.  In string theory (with unbroken supersymmetry), there are frequently axions.  These axions exhibit continuous shift symmetries in some approximation (e.g.\! perturbatively in the string coupling).  These are non-perturbatively broken, but often there is a discrete shift symmetry which is exact.  In other words, there is a dimensionless field, $a$, such that $a \rightarrow a + 2 \pi$ is an exact symmetry of the theory\footnote{In models exhibiting monodromy, this shift typically must be accompanied by transformations of other fields under which the vacuum state is not invariant.}.  In this formulation, $f_a$ depends on the precise form of the axion kinetic term.  The (non-compact) moduli which accompany these axions typically have Planck scale vev's (in a sense we will make precise in a moment).  The size of terms in the effective action associated with these fields is controlled by the $2 \pi$ periodicity of the {\it axions}.   If we denote the full chiral axion superfield by ${\cal A} = s + ia$,
this periodicity implies that, for large $s$, in the superpotential the axion appears as $e^{-n\cal A}$ for integer $n$.   In constructing models which include supersymmetry breaking, solving the strong CP problem requires suppressing only a small number of possible terms\cite{bobkovraby,dineaxions}.

Surveying known string compactifications, we might expect that there are several moduli which must be stabilized.  This is generally not a weak coupling problem.  In the framework of models with approximate supersymmetry (as above, this means supersymmetry
broken at scales well below the string scale),
the racetrack\cite{racetrack} idea and the KKLT model\cite{kklt} are scenarios for obtaining moduli stabilization in systematic approximations in a small parameter.  In both of these scenarios, the superpotential for the moduli plays an important role.  Whether indeed there are systems which realize these ideas with suitable parameters which can be taken arbitrarily small is unknown. 

But, in any case, one might think the axion multiplet is special.  If the superpotential plays a significant role in stabilization of the {\it saxion}, it is difficult to understand why the axion should be light.  More plausibly, the saxion might be stabilized by features of the \Kahler potential (\Kahler stabilization), in such a way that the imaginary part is not affected.  In perturbative string models, for example,
the \Kahler potential is often a function of ${\cal A} + {\cal A}^\dagger$.  There is no guarantee that would-be corrections to $K$ which stabilize ${\cal A}$ do not violate this symmetry substantially, but this is a widely adopted hypothesis.  

What would it mean to say that the dependence of the superpotential on $s$ is suppressed?  We might imagine that there is some other modulus, $T = t + ib$, appearing in the superpotential as $e^{-T}$, where $e^{-T}$ might set the scale for supersymmetry breaking.  We would also expect terms of order $e^{-n{\cal A}}, e^{-{\cal A}+T}$.  One possibility to account for the lightness of the axion would be that $s> t$, say $s=2t$.  Alternatively, $S$ and $T$ might be comparable, but $n>1$.  Then $T$ might be stabilized approximately
supersymmetrically in the manner discussed by KKLT:
\beq
W(T) = A e^{-T/b} + W_0
\eeq
with small $W_0$, leading to
\beq
T \approx b \log(W_0).
\eeq
 The potential for $s$ would arise from terms in the supergravity potential such as:
\beq
V_s = e^{K} \left \vert {\partial K \over \partial {\cal A}} W \right \vert^2 g^{{\cal A}~{\cal A}^*} + \dots
\eeq
For suitable $K({\cal A},{\cal A}^*)$, $V$ might exhibit a minimum as a function of $s$.    If $s$ is, say, twice $t$ at the minimum, $e^{-{\cal A}}$
is severely suppressed, as is the potential for the (QCD) axion, $a$. The fermionic component of ${\cal A}$
might be the goldstino (longitudinal component of the gravitino), but additional fields might well play a role.
In such a picture, the gravitino mass would be of order $e^{-t}$ up to powers of $t$ and $s$  (and possibly other fields), with the axion potential exponentially suppressed relative to $m_{3/2}$.

 Any such stabilization is inherently non-perturbative, and one might expect it to occur, generically, for small $s$.  On the other hand, the fact that gauge couplings are small in nature and the existence of hierarchies suggest that $e^{-s}$ is small (and similar factors for other moduli).  In particular, we might hope that an effective lagrangian analysis would be valid, in a Wilsonian action with cutoff scale well below $M_p$, and that we could organize the action in powers of $e^{-S},e^{-T}$.\footnote{It is quite possible -- even likely -- that there are situations where $e^{-S}$ is small in a region of the moduli space but the coefficients of $e^{-nS}$ are correspondingly large.  This is known to occur in some string world sheet analyses involving non-trivial compactifications. This would invalidate the underlying assumptions here in cases where it occurred.}.   For example, one might imagine that $e^{-t}$ accounts for the scale of supersymmetry breaking, while $e^{-t}$ accounts for the quality of the Peccei-Quinn symmetry.   We will the existence of such small exponentials for granted, as is rather standard in discussions of string phenomenology and cosmology, in this section, and discuss the issue in more detail in section \ref{modulieffectiveaction}.  

\subsection{A Remark on Distances in the Modulus Geometry}

Typical metrics for non-compact moduli fall off as powers of the field for large field.
Defining $s$ to be dimension one, 
\beq
g_{{\cal A},{\cal A}^*}=C^2 M_p^2/s^2
\eeq
for some constant, $C$.  So large $s$ is far away (a distance of order $C~M_p \log(s/M_p)$ away) in field space.  If, for example, the smallness of $e^{-(s + ia)}$ is to account for an axion mass small enough to solve the strong CP problem, we might require $s \sim 110~M_p$ or larger, corresponding to a distance of order $8 M_p$ from $s = M_p$ if $C = \sqrt{3}$.  

\section{Non-Compact Moduli as Inflatons}
\label{moduli}

We can summarize the previous section by saying that, if string theory and its (pseudo)moduli spaces are relevant to nature,  the strong CP problem points to Planck scale regions of field space as the arena for phenomenology.  Given this, it is important to consider non-compact moduli spaces as the setting for inflation.  As for the axion, we will assume some approximate supersymmetry.  As in ordinary, supersymmetric hybrid inflation, supersymmetry breaking is described by the $F$ component of the inflaton field (or a closely related field)\cite{dinehybrid2}.  One can write an effective field theory for the light fields (fields with masses of order $H_I$ or smaller) which is supersymmetric, with supersymmetry breaking described in terms of these supersymmetry breaking fields\cite{dinerandallthomas}.  This structure, indeed, explains why the moduli masses are typically of order $H_I$.  As is typical of such theories, tuning is required to account for the small values of the slow roll parameters (essentially why the inflaton is significantly lighter than $H_I$).  These issues will be discussed further in our subsequent discussion.

Our earlier discussion suggests some of the ingredients for such a structure:
\begin{enumerate}
\item  In the present epoch, one or more moduli responsible for hierarchical supersymmetry breaking.
\item  In the present epoch, a modulus whose superpotential is highly suppressed, whose compact component is the QCD axion.  This is not necessary for inflation, but is the essence of a modular (large field) solution to the strong CP problem.
\item  At an earlier epoch, a stationary point for some subset of fields in the effective action with higher scale supersymmetry breaking and a positive cosmological constant.  The current limits on gravitational radiation imply a suppression by at least $10^8$ relative to $M_p^4$.  Setting this aside, we might contemplate significantly lower scales.
\item  At an earlier epoch, a field with a particularly flat potential which is a candidate for slow roll.
\end{enumerate}
Fields need not play the same role in the inflationary era that they do now.  For example, the Peccei-Quinn symmetry might be badly broken
during inflation.  Then the axion will be heavy during this period and isocurvature fluctuations may not be an issue.  In such a case the initial axion misalignment, $\theta_0$, would be fixed rather than being a random variable.

Any would-be energy density is still eight orders of magnitude below $M_p^4$.  This suggests that moduli have large vev's, i.e. quantities like $e^{-{\cal A}},~e^{-T}$ we encountered before are quite small, though much larger than at present.  We might, for example, have a pair of moduli, ${\cal A}, T$ responsible for supersymmetry breaking, and an additional field, $I$, which will play the role of the inflaton.  During inflation, 
\beq
H_I \sim W \sim e^{-t}
\eeq
for example.  For typical \Kahler potentials, the curvature of the $t$ and $i$ potentials will be of order $H_I$.  We will exhibit a model with lower curvature below.  

A successful model of this sort requires a complicated interplay between effects due to the \Kahler potential and superpotential.  Corresponding to the general requirements we listed above, we need, for this set of degrees of freedom:
\begin{enumerate}
\item  The {\it potential} must possess at least two local, supersymmetry breaking minima in ${\cal A}$ and $T$, one of higher, one of lower, energy.  The former is the setting for the inflationary phase; the latter for the current, nearly Minkowski, universe.
\item  In the inflationary domain,
the potential for $I$, on the other hand, must be comparatively flat over some range.
\item  In the inflationary domain, the imaginary parts of ${\cal A}$ and $T$ should be comparable in mass to $H_I$ (or slightly larger) , if the system is to avoid difficulties with isocurvature fluctuations.  This would arise if $e^{-s} \approx e^{-t}$.
\item  In the present universe, the imaginary part of ${\cal A}$ should be quite light, along the lines described in the previous section.
The imaginary part of $I$ should be (as we will argue in more detail below) extremely light in the lower minimum.  
\item  There are additional constraints from the requirement that inflation ends.  For some value of Re $I$, the inflationary minimum for $T$ and ${\cal A}$ must be destabilized (presumably due to \Kahler potential couplings of $I$ to ${\cal A}$ and $T$).  At this point, the system must transit to another local minimum of the potential, with nearly vanishing cosmological constant.  
\item  The process of transiting from the inflationary region of the moduli space to the present day one is subject to serious constraints.  Even assuming that there is a path from the inflationary regime to the present one, the system is subject to the well-known concerns about moduli in the early universe\cite{bkn,brusteinsteinhardt}.  If they are sufficiently massive (as might be expected given current constraints on supersymmetric particles), they may reheat the universe to nucleosynthesis temperatures, avoiding the standard cosmological moduli problem.  $T$ and ${\cal A}$ are vulnerable to the moduli overshoot problem\cite{brusteinsteinhardt}, for which various solutions have been proposed.
\end{enumerate}

\section{Effective Theories for Modular Inflation}
 \label{modulieffectiveaction}

In the previous section, we have listed some requirements for a successful theory of modular inflation.  Needless to say, we don't know how to extract such a theory from string theory.    In this section, we will content ourselves with writing down theories which satisfy (some of) the various requirements separately.  We first discuss the question of the validity of effective field theory in the regimes of interest.  Then we turn to features of the \Kahler potential and superpotential necessary to satisfy the requirements of modular inflation.
 
 \subsection{The effective action on the moduli space}

 In any possible theory of quantum gravity, the effective action has, at best, limited applicability.
 Moduli stabilization as described above occurs, more or less by definition, in a region where standard perturbative methods are not valid\cite{dineseibergstrongcoupling}.  In particular, in the large field regime in string theory, the metric for the moduli tends uniformly to zero and does not exhibit interesting structure.  In cosmological situations, the motion on the moduli space exhibits singularities\cite{banksdinesingularities}.  Finally, there are more general reasons of principle to question the use of effective field theory methods\cite{bankseft}.
 
 If moduli are stabilized in a region where weak coupling methods are not valid, in what sense might an effective action be useful or even meaningful?  Without solving a (non-supersymmetric) string theory, this is a hard question to answer, but we can at least formulate a set of underlying -- and widely held -- assumptions which would provide a rationale for such a treatment.  It is easiest to articulate the underlying assumptions -- and some of their limitations -- if we consider a moduli space which asymptotically is approximately supersymmetric.
 
 The standard assumptions of string phenomenology are that moduli are stabilized in regions where the observed gauge couplings are weak; usually it is also assumed that there is an approximate supersymmetry with a hierarchy of scales.  Suppose, first, that this occurs due purely to high energy effects, without, for example, the action of a strongly interacting gauge theory.  Then the effective low energy theory might consist of moduli only (plus supergravity).  Ignoring gravity at first, we can argue self consistently for the validity of a low energy effective theory, in much the way one argues for the non-linear pion lagrangian.  In particular, we can study slowly varying classical field configurations.  The existence of a light gravitino in regions of the field space implies that the effective lagrangian (in these regions) will be supersymmetric (with supersymmetry breaking described within the action itself).  The fields can then be organized into chiral multiplets; the superpotential is constrained by the axion shift symmetries.  This implies that for a range of fields, the superpotential is exponentially small\footnote{Though it might be the sum of exponentially small terms with large coefficients}.  It is difficult to make arguments about the \Kahler potential.  Indeed, it has been argued\cite{kahlerstabilization} that large corrections to the \Kahler potential might be the source of moduli stabilization and
we have already invoked this possibility.  
 
Now including gravity, the effective action has, at best, limited validity.
For effective actions involving moduli and gravity, apart from stable AdS and Minkowski space, typical 
motions on the space will begin or end in singularities, where necessarily the effective field theory treatment breaks down\cite{banksdinesingularities}.  The understanding/resolution of such singularities may require a framework such as eternal inflation or something yet unknown.  In addition, the moduli space itself is expected to have singularities.
For example, in a one complex dimensional space, with a constant superpotential, if the metric behaves asymptotically as $1/(S + S^\dagger)^2$, there cannot be a stationary point unless the metric vanishes somewhere\cite{kahlerstabilization}\footnote{The issue of the vanishing metric was raised by Kaplunovsky in a private communication}.  One might expect these sorts of singularities even if there are multiple moduli, but they may be isolated, and there may be (and we will assume there are) cosmological histories which avoid them.  They might be connected with the appearance of new massless states, or some strong coupling (or both).

So we content ourselves with the assumption that for the period of cosmic history of interest (inflation up to the present time) the motion of the system be through non-singular regions of the space.  We require our effective description only be valid during this period\footnote{An attempt to formulate a more global picture of cosmology in quantum gravity is the holographic cosmology of \cite{banksfischler}}.

Such smooth behavior is indeed a requirement for inflation; if the metric on the space of fields becomes singular, fields become strongly coupled and it is hard to imagine that the slow roll conditions can be satisfied.   It would seem to be a feature of the more recent history of the universe as well.   For the discussion of inflation, it is necessary that there be trajectories which avoid the singular regions.   
 To summarize, we assume that for the relevant period of cosmic history, motion on the pseudomoduli space can be described by a non-singular effective field theory.  This does not mean that the moduli space is non-singular everywhere, nor that the full cosmic history is smooth (or that this history is everywhere describable through this effective theory).
 
\subsection{Stabilizing Moduli in the Current Universe}

We have already discussed stabilization of the moduli ${\cal A}$ and $T$ within the context of large field solutions of the strong CP problem.  We implemented stabilization of $T$ in a manner similar to that of KKLT.  We required that the terms in ${\cal A}$ in the superpotential be exponentially suppressed relative to $e^{-T/b}$, so that $s$ was stabilized by \Kahler potential effects, while the leading contributions to the $a$ potential arise from QCD.  We require, for a successful inflationary model, that ${\rm Re}~I$ also be fixed by \Kahler potential effects.  The goldstino is a linear combination of $I$ and ${\cal A}$.  The imaginary part of $I$ must be extremely light, so it does not constitute an appreciable part of the energy density today.

\subsection{Stabilizing Moduli During Inflation}

During inflation, we require $e^{-T}$ be small, but much larger than its value in the present universe.  This might be achieved, in an approximately supersymmetric fashion, with a superpotential of the form:
\beq
W = \epsilon e^{-T} + e^{-2T} + e^{-(T + {\cal A})}+ e^{-{\cal A}} +W_0.
\eeq
Here $\epsilon \sim 10^{-2}$, and $W_0$ is far smaller.  Then there is a local minimum at $e^{-T}, e^{-\cal A} \sim \epsilon$.  At this minimum, the masses of all of the components of $T,{\cal A}$, and in particular, the axion, are comparable and can readily be somewhat large compared to the scale of inflation (by powers of $T$, $s$).

$I$ should not appear in the superpotential (or should be further suppressed).  Sufficient flatness of the $I$ potential to account for inflation places severe restrictions on the \Kahler potential.  We will discuss models for the \Kahler potential and the inflaton potential in the next section.

\subsection{Requirements for the Transition Period}

Couplings of the inflaton to the fields $S$ and $T$ must destabilize the inflationary minimum and end inflation.  Given the restrictions we are imposing on the superpotential, this requires that couplings in the \Kahler potential, such as 
\beq
(I+I^\dagger)^2 ((S+S^\dagger)^2,(T+T^\dagger)^2)
\eeq
contribute negatively to the $s$ and $t$ masses at the later stages of inflation, in such a way that the system can transit smoothly from one state to the other.  The details of the terms in the action which dominate during this period are likely to be
rather involved, with mixings, for example, among the various fields during this phase.  During the inflationary phase itself the requirement is that $K$ is such that $I$ is much lighter than the other fields (whose masses are of order $H_I$, the 
Hubble scale during inflation, or larger).  One can contemplate different possibles for
the masses of these fields and their stabilization in the present universe.  Given that, for this discussion, the Kahler potential
is an arbitrary function of the fields, we do not see any difficulty in principle with satisfying these conditions.  In the next section, we will be more explicit about possible behaviors of $K$ and $W$ during the inflationary era.

It should be stressed that the requirements for destabilization are, like those for stabilization, not consistent with a systematic expansion in $1/s$, $1/t$.  In the region of large $s,t$, in particular, one would expect these fields to have masses parametrically large compared to $m_{3/2}$, and the \Kahler couplings could not destabilize this minimum.  We should stress again, as well,
that typical Kahler potentials may lead to difficulties in settling into the correct, final vacuum\cite{brusteinsteinhardt}.
The Kahler potential may have to be somewhat singular.  This question will be explored elsewhere.

\section{Inflationary Models:  Large $r$}

We consider, first, as a benchmark, the case where $r$ is in the range of the reported BICEP2 result, corresponding to field excursions several times $M_p$.  In this regime, it is easy to construct models consistent with the data on $r$ and $n_s$ as well as the fluctuation spectrum.
As an example, we take the \Kahler potential for $I$ to be:
\beq
K = -{\cal N}^2 \log(I + I^*).
\eeq
Writing
\beq
I = e^{\phi/{\cal N}}
\eeq
the kinetic term for $\phi$ is simply $(\partial \phi)^2$.  ${\cal N}$ will serve as the small parameter accounting for slow roll.  In the limit of very small ${\cal N}$, the potential for $\phi$ becomes flat.

Indeed, the $\phi$ potential, with the assumption that all moduli besides $I$ are fixed (and have masses greater than $H$), and
that $I$ does not appear in the superpotential, with the Kahler potential above is
\beq
V(\phi) = e^{-{\cal N} \phi} V_0,
\eeq
$V_0$ being the minimum of the $S$, $T$ potential.
So the slow roll parameters are:
\beq
\epsilon = {1 \over 2} {\cal N}^2; \eta = {\cal N}^2 = 2 \epsilon.
\eeq
Note
\beq
n-1 = -2 \epsilon.
\eeq
If $n_s = 0.96$, as measured by the Planck satellite\cite{planckinflation},
 $\epsilon= - 0.02$, and $r = 16 \epsilon = 0.32$.  This is well above the recent quoted limit on $r$
 from the joint BICEP2/Planck analysis\cite{bicep2planck}.
In section \ref{smallr}, we will discuss modifications which, with tuning, permit smaller values of $r$.  From this viewpoint,
one would expect $r$ to be as large as permitted by present observations.

A model with features similar to that of this section (with $\cosh$ rather than exponential potential) has been discussed in \cite{nojiri}.  Ref. \cite{burgesslargefield} discusses a number of issues in large field inflation, and writes models
with features similar to those of this section.

\subsection{Connection to Chaotic Inflation}

Chaotic inflation\cite{chaotic} has, for decades, provided a simple model for slow roll inflation, and its prediction of transplanckian field motion and observable gravitational radiation now may be validated.  As we look at the moduli inflation model of the previous section (and more generally moduli models of large field inflation), we see, in fact, a realization of the ideas of chaotic inflation.  Again, the potential behaves as
\beq
V \sim H_I^2 M_p^2 e^{-{\cal N} \phi}
\eeq
We have seen that the exponent changes, during inflation, by a factor of about $3/2$.  So we can make a crude approximation, expanding the exponent and keeping only a few terms.  If we focus on each monomial in the expansion, the coefficient of 
$\phi^p$, in Planck units, is:
\beq
\lambda^p = {10^{-8} {\cal N}^p \over p!},
\eeq
where $N$ is the number of $e$-foldings.
We can compare this with the required coefficients of chaotic inflation driven by a monomial potential, $\phi^p$.  In this case,
\beq
\lambda_p = {3 \times 10^{-7} p^2 \over (2Np)^{{p \over 2} + 1}}
\eeq
These coefficients are not so different.  For example, for $p=1$, the moduli coefficient is about $2 \times 10^{-9}$, while for the chaotic case it is about four times smaller; the discrepancy is about a factor of two larger for $p=2$.  So we see that these numbers, which would one hardly expect to be identical, are in a similar ballpark. 

So moduli inflation provides a rationale for the effective field theories of chaotic inflation. The typical potential is not a monomial, but one has motion on a non-compact field space, over distances of several $M_p$, with a scale, in Planck units, roughly that expected for chaotic inflation.  The structure is enforced by supersymmetry and discrete shift symmetries.

\section{Moduli Inflation:  Small $r$}
\label{smallr}

It is now clear that $r < 0.1$\cite{bicep2planck}, and it is conceivable that it is significantly smaller.  Still, as we explain in this section, moduli inflation provides a setting, where fields would be of order $M_p$, but their excursions would be small (as we will see, of order $M_p$ or less).  An interesting question is whether these models are more or less tuned than models with a higher scale.  We will argue, in fact, that the lower scale models {\it are} more highly tuned; this might be an argument in favor of high scale inflation.

In the large $r$ model of the previous section, $\epsilon$ and $\eta$ were naturally comparable.  In small $r$ models, given our knowledge of $n_s$, we have $\epsilon \ll \eta$.  We can be rather explicit if we assume that the superpotential (as in the previous models) is roughly constant during inflation ($W_I$), and similarly that the order parameter for supersymmetry breaking, $F_Z$, is roughly constant.
Then
\beq
V = e^{K}\left [ \left \vert {\partial K \over \partial I} W_I \right \vert^2 g^{I I^*}  + V_0.  \right ].
\eeq 
 Here $V_0$ is a combination of the supersymmetry breaking terms and $-3 \vert W_I \vert^2$.  
 
 In the moduli cosmology framework, an inflationary model is characterized by a choice of \Kahler potential and superpotential.
The requirements that $V$ produce the desired values of $\epsilon$ and $\eta$ yield constraints on $K$.  In general, smaller $\epsilon$ implies greater tuning of the \Kahler potential.  The Planck results for $n_s$, in addition, require that $\eta$ be negative.  To demonstrate that these requirements can be met, we study a class of \Kahler potentials:
\beq
K(I,I^*) = -{\cal N}^2 \log(I + I^*) + {A \over I + I^*} + {B \over (I + I^*)^2}.
\eeq
We treat ${\cal N}$ as fixed, as well as $I$.  We then take $B = B(A)$ such that $n_s = 0.96$, and vary $A$.
In order to achieve small $\epsilon$ we demand that we sit near a point where $V^\prime \equiv \partial V/\partial \phi = 0$.  At such a point,
$\epsilon = {d\epsilon \over dA} = 0.$  So
\beq
\epsilon = {1 \over 2} {d^2 \epsilon \over dA^2} \delta A^2.
\eeq
We see, in this way, that studying just this subset of parameters, the tuning in $A$ goes as $\sqrt{\epsilon}$.  Considering the full set of parameters, the tuning is arguably somewhat more severe.

We indeed find that for a range of parameters, one can have suitable $n_s$ with small $\epsilon$.  As an example of a point with $V^\prime =0$,
\beq
{\cal N}= .0256 ;~i =7.70304;~A= .00048800;~B=0.000554.
\eeq
So, for example, if $V= (10^{14} ~{\rm GeV})^4$,
then
$A =. 00048795$, indicating the required degree of tuning, about a part in $10^5$, as expected.


To summarize this section, small field inflation can be achieved within the moduli models.  The fields should be thought of as Planck scale, but the excursions are Planck scale or less.  These models are more tuned than the higher scale models, so this is an a priori argument in favor of higher scale inflation.

\section{Conclusions}

Explaining inflation from an underlying microscopic theory is an extremely challenging problem, quite possibly inaccessible to our current theoretical technologies.  As we have reviewed, even in so-called small field inflation, it requires control over Planck scale phenomena.  Within string theory, this requires understanding of supersymmetry breaking (whether large or small) and fixing of moduli in the present universe as well as at much earlier times.  It requires an understanding of cosmological singularities, and almost certainly of something like a landscape.

That said, as we reviewed, the essence of hybrid inflation is motion on a (non-compact) pseudomoduli space.  In string theory, at least at the classical level, such moduli spaces are ubiquitous, and the features of these moduli suggest a picture for inflation in which the (canonical) fields have Planck scale motions.  
We have stressed a parallel between small/large field inflation and small/large field solutions to the strong CP problem.  The existence of moduli in string models is strongly suggestive of the large field solutions to both problems.  The proposal we have put forward here is similar to the large field solutions of the strong CP problem.

We have noted that in such a picture, several moduli likely play a role in inflation, achieving the needed degree of supersymmetry breaking and slow role.  We have seen that small $r$ is more tuned than large $r$, giving some weight to the former possibility.  We have stressed the contrast with small field inflation, where extreme tuning to achieve low scale inflation is replaced by the requirement of an extremely small dimensionless coupling.  

Returning to the strong CP problem, we have stressed that any would-be Peccei-Quinn symmetry is an accident, and that the accident which holds in the current configuration of the universe need not hold during inflation; this would resolve the axion isocurvature problem.

We conclude by asking, in such a framework, what one means by a {\it model} and what might be tests or predictions of the framework.  More precisely, the inflationary paradigm is highly successful; the question is whether we can provide some compelling microscopic framework and whether it is testable.
As always in inflationary model building, this is a difficult question.  First, one is typically introducing a great
deal of structure with which to explain a small number of parameters.  One has additional fields, which may, in the present universe, be quite heavy, and also may be weakly interacting with Standard Model particles.  Moduli inflation suffers from the same issues.  Indeed, our point of view has not been to obtain a particular, detailed model,
but to ask what features might be generic in a plausible scenario for the implementation of
inflation in the sorts of effective field theories which seem to emerge from string theory (or quantum
theories of general relativity more generally). Two features which would seem generic would be:
\begin{enumerate}
\item   Higher scales of inflation are preferred.
\item   High scale axions (even an {\it axiverse}\cite{axiverse}).
\end{enumerate}
In a more detailed picture, one might hope to connect some lower energy phenomenon, such as supersymmetry breaking, with
inflation.  This would require significant more work to embed these models in theories with supersymmetry breaking, and to understand in more detail the end of inflation and the transition to a state more closely resembling our current universe.

\vspace{1cm}

\noindent
{\bf Acknowledgements:}  This work was supported by the U.S. Department of Energy grant number DE-FG02-04ER41286.  We thank Tom Banks and Patrick Draper for discussions.

 \newpage

\bibliographystyle{unsrt}
\bibliography{inflation_after_biceps_revised_september_2015}{}

\end{document}